\documentclass[12pt]{article}

\author{Vladimir A. Petrov \footnote{e-mail: Vladimir.Petrov@ihep.ru}}

\title{Bethe's Ansatz for Coulomb -Nuclear Interference Is Incompatible with Additive Eikonal}
\date{}

\begin{document}

\maketitle
\begin{center}
A. A. Logunov Institute for High Energy Physics 

NRC "Kurchatov Institute", Protvino, RF
\end{center}

\begin{abstract}
This is a proof that if the eikonal is, as usually assumed, additive in strong and electro-magnetic interactions then the application of the Bethe Ansatz for the full scattering amplitude leads to the strong interaction scattering amplitude with the ratio ( real/imaginary) independent on the transferred momentum $ t $. Moreover, the unitarity condition makes the strong interaction amplitude vanishing.

Thus, the Bethe form for the Coulomb-nuclear scattering amplitude and the same amplitude based on the additive eikonal are incompatible.
\end{abstract}

\section{Introduction}	

Recent measurements by the TOTEM at 13 TeV caused a vivid discussion (more than 60 publications by now) of a strikingly small value of the parameter $ \rho = {Re T(s,0)}/{Im T(s,0)} $ which lies (with some variations) near $ 0.10 $. The extraction of this parameter (which, unfortunately, is inherently model dependent) relies heavily on how the Coulomb contributions leading to Coulomb-nuclear interference (CNI) are taken into account in the full scattering amplitude.

Since the pioneering paper by H.Bethe \cite{Be} it became a long-standing practice to use the following form for the full (including both strong (N) and Coulomb (C) interactions) scattering amplitude:
\begin{equation}
T_{C+N} (s,t) = T_{C} e^{i\alpha\varphi(s,t)}+ T_{N} (s,t)
\end{equation}
where
\begin{center}
$T_{C} = 8\pi s \alpha F^{2}(t)/t  $
\end{center}
for $ p p $ scattering with $ F(t) $ the proton form factor.

There were suggested many expressions for the "West-Yennie phase" $ \varphi(s,t) $ (to cite a few, see \cite{We}) which contain both electromagnetic and strong interaction parameters, e.g., the forward slope
$ B (s) $, the proton "charge radius" etc. which allowed description of the data of various degrees of quality. 

The very Bethe Ansatz (1) is not at all a rigorously and generally established  formula. Early criticism of the soundness of the Bethe Ansatz (1) in the form
 given by West and Yennie \cite{We} can be found in Ref.\cite{Sol} where, in particular, it was shown that at relativistic energies the "West-Yennie phase" $ \varphi (s,t) $, which has to be real by definition,  acquires an imaginary part.
 At low and intermediate energies the Bethe parametrization was proven unsatisfactory \cite{Li} from phenomenological reasons.
 
 Later on it was also noticed that to keep the Bethe ("West-Yennie") phase real one should assume that the nuclear amplitude would have a  $ t $-independent phase and therefore a new, more general formula was suggested in Refs.\cite{Ku}, \cite{Ca}. In Ref.\cite {Pe} the new arguments were presented against the use of the "Bethe Ansatz" in the case of point like proton charges and a modification of the general formula derived in Refs.\cite{Ku}, \cite{Ca} for the smeared charge case has been given.
 
 In a sense, the present paper can be considered as a conceptual completion of the previous criticism from the additive eikonal viewpoint. I will present a proof of the inconsistency of the "Bethe Ansatz" in general case if to remain with the  basic Bethe's assumption about additivity of the eikonal in electromagnetic and strong interactions (see Eqs.(4.28)-(4.30) in Ref.\cite{Be}). 
 
  In truth, Bethe himself, who acted in the framework of a particular potential model, did not seem to claim the universal applicability of his parametrization, but later it was "canonized" and we treat it here that way.
  In no way, however, the Bethe or the additive eikonal form for the CN scattering amplitude  has the advantage of one over the other: they are just incompatible.
\section{The general framework}
To forestall possible misunderstanding I give in this Section a cursory survey of the present state of knowledge concerning the problem of account of the soft photon radiation
(both real and virtual)in scattering of charged particles.
As is well known the (see, e.g. \cite{Akh}) differential cross-section of "elastic" scattering of charged hadrons
is actually an inclusive cross-section for the process (for definiteness we'll consider the proton-proton case)
\begin{center}
$ p+p \rightarrow p+p+X(soft\, photons)  $
\end{center} and, accordingly, the experimentally observable cross-section is 
\begin{equation}
\frac{d\sigma^{C+N}_{obs}}{dt} = \int_{0}^{\bar{M}^{2}}dM^{2}\frac{d\sigma^{p+p \rightarrow p+p+X(soft\, photons) } }{dt dM^{2}} 
\end{equation}
\[= \exp ( - \Gamma (s,t,\bar{M}^{2})\frac{d\sigma_{C+N}}{dt}\]
where 
\begin{equation}
\frac{d\sigma_{C+N}}{dt}= \frac{(\hbar c)^{2}}{16\pi s^{2}} \mid T_{C+N}\mid^{2} 
\end{equation}
while $ \bar{M}^{2} $ is the upper limit for the missing mass of the undetectable soft radiation related to the momentum accuracy of the protons, the damping factor $ \exp ( - \Gamma (s,t,\bar{M}^{2}) $ being resulted from interplay of real and virtual (soft) photons and making the cross-section decreasing with the transferred momenta growth. 
In Eq.(2) $ t $ means some kind of average of two momenta $ t_{1,2} $ related to the inclusive process and $ t_{1,2} \sim t \pm \mathcal{O}(\bar{M}^{2})$ with $\bar{M}$ of order on uncertainty in moments of interacting protons.

At $ -t\rightarrow 0 $ the damping gets negligible.
Eq.(2)is normally being derived in the S-matrix approach with IR regulators (say, in the form of fictitious "photon mass") which are afterwards removed when taking physical probabilities( cross-sections)in which IR singularities are cancelled.
More direct (though physically absolutely identical to the standard one) approach dealing with massless photons from the very beginning and using instead of usual one particle in- and out states (inadequate in presence of massless fields) the coherent states of "dressed charges" was developed in Refs.\cite{Kul} but it still did not find wide use in phenomenological applications.

In what follows we will deal with $ d\sigma_{C+N}/dt $ which is the proton-proton "elastic" scattering cross-section accounting for only strong and Coulomb exchanges between the colliding protons. At small enough $ t (u) $ it  practically coincides with the observed cross-section (2).

As follows from the eikonal representation of the full (C+N) scattering amplitude 
 $ T_{C+N} $, its modulus  can be cast into the form \cite {Pe} 
\begin{equation}
\mid T_{C+N}(s,t) \mid_{t,u \neq 0} \: = \: \mid -2is\hat{\Xi}^{\alpha}(q)+ \int\frac{d^{2}q^{'}}{(2\pi)^{2}}\hat{\Xi}^{\alpha}(q-q^{'})T_{N}(s,t^{'}) \mid
\end{equation}
where
\begin{center}
 $\hat{\Xi}^{\alpha}(q) = \int d^{2}b e^{iqb} e^{2i\hat{\delta}_{C}(b)}$,
\end{center}
and Coulomb eikonal reads

\begin{center}
$ \hat{\delta}_{C}(b)= \frac{1}{4s} \int d^{2}q/(2\pi)^{2} [exp(-iqb) -1]T_{C}(q^{2}) =$ 
\end{center}

\begin{center}
$ = -\alpha \int_{0}^{\infty}\frac{d\mid q\mid}{\mid q\mid} F^{2}(q^{2})[J_{0}(\mid q\mid \mid b\mid)-1] $
\end{center}

while 
\begin{center}
$  T_{N}(s,t) = 2is \int d^{2}b \exp (iqb) (1 - \exp (2i\delta_{N}(s,b)) $
\end{center}
is the strong interaction amplitude induced by the eikonal $ \delta_{N}(s,b) $. 
Notice that the amplitude \emph{modulus} if free of non integrable singularities due to the masslessness of the photon as it surely should be.

Eq.(4) stems from the commonly assumed additivity of the full eikonal:
\begin{center}
$\delta_{C+N}(s,b)= \delta_{N}(s,b)+\delta_{C}(s,b)  .$

\end{center}
In what follows we will use along with $ t(u) $ the variable \begin{center}
$  q^{2} \equiv q^{2}_{\perp} = ut/4k^{2} = k^{2}sin^{2}\theta, \; s= 4k^{2}+ 4m^{2},  $
\end{center}
which reflects the $ t-u $ symmetry of the $ pp $ scattering.
At $ \theta\rightarrow 0\;\quad q^{2} \approx -t $ while at $ \theta\rightarrow \pi\;\;\quad q^{2} \approx -u $.

In order not to lose generality, we do not refer to any specific model for the amplitude $ T_{N}(s,t) $ (or the eikonal $ \delta_{N}(s,b) $) of strong interaction, with the exception, perhaps, of a general property to rapidly drop at large $ q^{2} $, analyticity at $q^{2}=0 $ and unitarity.

\section{“When in doubt, expand in a power series.” (E. Fermi)}
 
 What we are going to do next is to expand two expressions
 \begin{equation}
 \mid T_{C} e^{i\alpha\varphi(s,t)}+ T_{N} (s,t)\mid 
 \end{equation} 
 and 
 \begin{equation}
 \mid -2is\hat{\Xi}^{\alpha}(q)+ \int\frac{d^{2}q^{'}}{(2\pi)^{2}}\hat{\Xi}^{\alpha}(q-q^{'})T_{N}(s,t^{'}) \mid
 \end{equation}
 which tentatively represent one and the same quantity, i.e. $ \mid T_{C+N}(s,t) \mid $,
 in powers of $ \alpha $ and to compare the corresponding coefficients. This should give us possible restrictions on the strong interaction amplitude which the  Bethe Ansatz implies (5) (if any).
 
 Let us notice first that the expansion of a modulus $ \mid f(\alpha)\mid $ in $ \alpha $
 up to the first order is of the form
 \begin{center}
 $\mid f(\alpha)\mid = \mid f(\alpha=0)\mid + \lbrace Re[f^{\ast}(\alpha=0)\cdot f^{'}(\alpha=0)]/ \mid f(\alpha=0)\mid\rbrace \cdot \alpha + \mathcal{O}(\alpha^{2}) $
 \end{center}
 where $ f^{'}(\alpha) \equiv \partial f/\partial \alpha $.
 The coefficient $\lbrace Bethe \rbrace$ before $ \alpha $ which stems from the Bethe
 Ansatz (3) reads:
 \begin{equation}
 \lbrace Bethe \rbrace = F_{C} ReT_{N}/\vert T_{N}\vert
 \end{equation}
 while the general expression (4) gives
 \begin{center}
 $\lbrace General \rbrace =  F_{C} ReT_{N}/\vert T_{N}\vert - $
 \end{center}
 \begin{equation}
 - \frac{1}{2s \vert T_{N}\vert}\int\frac{d^{2}q^{'}}{(2\pi)^{2}}F_{C}(q^{'}) \lbrace ReT_{N}(q)ImT_{N}(q-q^{'})-ImT_{N}(q)ReT_{N}(q-q^{'})\rbrace,
 \end{equation}
here\begin{center}
$ F_{C}(q) = -8\pi s F^{2}(q^{2})/q^{2} $.
\end{center} 
Comparing expressions $\lbrace Bethe \rbrace$ and  $\lbrace General \rbrace$ we come to

\begin{center}
$ \int\frac{d^{2}q^{'}}{(2\pi)^{2}}F_{C}(q^{'}) \lbrace ReT_{N}(q)ImT_{N}(q-q^{'})-ImT_{N}(q)ReT_{N}(q-q^{'})\rbrace = 0 . $
\end{center}

 As this equation holds for arbitrary $ q-q^{'}  $ we get
 \begin{center}
 $ ReT_{N}(q)ImT_{N}(q^{'})-ImT_{N}(q)ReT_{N}(q^{'})=0 , \;\forall q, q^{'}.$
 \end{center}
 We have thus that $\textit{ Arg }T_{N}(s,t) $ does not depend on $ t $. Thus, the scattering amplitude has to have (according to the Bethe formula (1)) the following general form\begin{equation}
 T_{N}(s,t)= T (s+i0,t)\vert _{s\in r.h. cut}= (1+i\omega(s))ReT_{N}(s,t),
 \end{equation}
 the function  $ \omega $ is simply related to the familiar parameter $ \rho $:
 \begin{center}
 $ \omega(s) = 1/\rho(s), \:\omega(s\:\bar{\in}\:cuts) = 0  .$
 \end{center}
  The quantity $ ReT_{N}(s,t) $ is a distribution in $ s $ and an analytic function in $ t $. 
 
 To check the acceptability of such an option for the scattering amplitude let us take the energy lying below the inelastic threshold. It follows immediately from Eq. (9) and elastic unitarity ( $ T_{l}(s) =r_{l}(s) + i a_{l}(s)  $ )
 
 \begin{center}
 $  a_{l}(s) = a_{l}^{2}(s) + r_{l}^{2}(s), \; s_{el} \leq  s  < s_{inel}, $
 \end{center}
 that the partial amplitude
 \begin{center}
 $  r_{l}(s) = \frac{1}{32\pi \sqrt{s}k} \int^{0}_{-4k^{2}} dt P_{l}(1+t/2k^{2}) ReT_{N}(s,t)$ 
 \end{center}
  turns out to be independent on $ l $
 \begin{equation}
 r_{l}= \omega (s)/(1+\omega^{2}(s))= \rho(s)/(1+\rho^{2}(s)).
 \end{equation}
 This results in a discouraging outcome:
 \begin{center}
 $ T_{N}(s,t)= (i + \rho(s))/(1+ \rho^{2}(s))[8k^{2}\delta (t)] = 0 , \;s_{el} \leq  s  < s_{inel}, $
 \end{center}
 if to recall our condition in Eq.(4) ($ t\neq 0 $). 
 
In this way the comparison of the eikonal representation and the Bethe formula leads to a flagrant contradiction with analyticity of the strong interaction scattering amplitude at $t=0  $ while the very amplitude vanishes at $ t\neq 0 $ and energies below the first inelastic threshold. As the physical amplitude in question is a  value of the analytic function of $ s $ on boundary of the analyticity domain, we see that actually the cut between elastic and inelastic thresholds is absent as the amplitude is zero there. 
We see that an analytic function vanishes on a finite segment inside the analyticity domain. From the uniqueness of analytic functions it follows that the whole amplitude vanishes.

 \section{ Conclusion and outlook}
 
\textit{Thus we arrive to the conclusion that the Bethe Ansatz (1) leads , in the case of an additive eikonal, to a vanishing strong interaction amplitude.}

 Such a disappointing conclusion still leaves a room for further insights. Actually the additivity of the eikonal came from the non relativistic context where potentials of different forces just add. In relativistic case we have no proof of the corresponding factorization of the elastic scattering matrix 
 \begin{center}
 $ S_{CN}(b) =S_{C} (b) \cdot S_{N} (b) $
 \end{center}
 which embodies the eikonal additivity. 
 
 Generally, we can always represent the full eikonal in the form
 \begin{center}
 $  \delta_{CN}(b, \alpha) + \delta_{N}(b) $
 \end{center}
 where $ \delta_{CN}(b, \alpha) $ is an "irreducible" part of the eikonal which depends both on strong interaction and electromagnetic parameters  but vanishes when the electromagnetism is switched off ( $ \delta_{CN}(b, \alpha=0) =0 $).
 It would be interesting to see if, in this general case, the Bethe Ansatz leads to acceptable restrictions, if any,  on the strong interaction amplitude.
 
 Our conclusion is: both Bethe Ansatz and the \textit{additive} eikonal representation for the Coulomb-nuclear scattering amplitude which are based on specific assumptions are neither proved nor ruled out by the present analysis. We only prove that they are incompatible between each other. We believe that awareness of this fact may be useful in phenomenological studies.

 \section{Acknowledgements}

 I am grateful to  Vladimir Ezhela, Anatolii Likhoded, Andr\'{e} Martin, Roman Ryutin and Nikolai Tkachenko
 for their interest to the subject and inspiring conversations. I am  particularly indebted to Anatolii Samokhin for very fruitful discussions.

 This work is supported by the RFBR Grant 17-02-00120.

\end{document}